%% file: main.tex
\documentclass[journal]{IEEEtran}
\IEEEoverridecommandlockouts
\usepackage{cite}
\usepackage{amsmath,amssymb,amsfonts}
\usepackage{algorithmic}
\usepackage{graphicx}
\usepackage{textcomp}
\usepackage{tikz}
\usetikzlibrary{patterns}
\usepackage{xcolor}
\def\BibTeX{{\rm B\kern-.05em{\sc i\kern-.025em b}\kern-.08em
    T\kern-.1667em\lower.7ex\hbox{E}\kern-.125emX}}

\usepackage{stfloats}

\usepackage{amsthm}

\tikzset{every picture/.style={line width=0.75pt}} %set default line width to 0.75pt

\DeclareMathOperator*{\argmax}{arg\,max}
\DeclareMathOperator*{\argmin}{arg\,min}
\begin{document}

\title{Band \& Tone Jamming Analysis and Detection on LoRa signals}

\author{\IEEEauthorblockN{Cl\'ement Demeslay\IEEEauthorrefmark{1}, \textit{Student Member, IEEE},
Roland Gautier\IEEEauthorrefmark{2}, \textit{Member, IEEE}, Anthony Fiche \IEEEauthorrefmark{3}, \textit{Member, IEEE} and Gilles Burel\IEEEauthorrefmark{4}, \textit{Senior Member, IEEE} \\}
\IEEEauthorblockA{Univ Brest, Lab-STICC, CNRS, UMR 6285, F-29200, France\\
Email: \IEEEauthorrefmark{1}clement.demeslay@univ-brest.fr,
\IEEEauthorrefmark{2}roland.gautier@univ-brest.fr,
\IEEEauthorrefmark{3}anthony.fiche@univ-brest.fr}, \IEEEauthorrefmark{4}gilles.burel@univ-brest.fr}

\maketitle

\begin{abstract}
This paper examines the effect of Band Jamming ($BJ$) and Tone Jamming ($TJ$) on LoRa signals in a flat Additive White Gaussian Noise ($AWGN$) channel.
In this scenario, LoRa proves to have good resiliency against these jamming attacks.
Furthermore, a simple and lightweight $BJ$ and $TJ$ jammer detection scheme is derived.
Theoretical and simulation results show good detection capability, especially with Single Tone Jamming ($STJ$).
\end{abstract}


\section{Introduction}

The Internet of Things ($IoT$) is experiencing striking growth since the past few years enabling much more devices to communicate and allowing many scenarios to be a reality such as smart cities.
The number of $IoT$ devices is expected to rapidly grow, jumping from almost 10 to more than 21 billion \cite{IoTDevicesNumber}.
Many technologies were developed in that sense relying on licensed bands (Narrow Band $IoT$ ($NB-IoT$), Extended Coverage GSM ($EC-GSM$) and LTE-Machine ($LTE-M$)) or unlicensed bands such as SigFox, Ingenu, Weightless or Long Range (LoRa) \cite{goursaud}.
This paper focuses on LoRa standard.
LoRa has been initially developed by the French company Cycleo in 2012 and is now the property of Semtech company, the founder of LoRa Alliance. 
LoRa is nowadays a front runner in LP-WAN solutions and holds a lot of attention by the scientific research community. 
% Due to its patented nature, initial research was mainly based on retro-engineering of existing LoRa transceivers \cite{knight_gnuradio}.
% The first paper to provide a rigorous mathematical representation of LoRa signals and its demodulation scheme was achieved by L. Vangelista in \cite{vangelista}. 
% Many research was conducted focusing on LoRa network capacity enhancements \cite{joerg3}, channel coding improvements \cite{joerg4}, temporal and frequency synchronization techniques \cite{xhonneux} or practical approaches to decode a LoRa symbol contaminated by a single or multiple LoRa users \cite{laporte}.
The LoRa vulnerabilities were addressed in the literature.
In \cite{huang}, a malicious LoRa user acts as a reactive jammer by sending random LoRa symbols to a legitimate LoRa node.
The authors evaluated the jamming impact on Packet Delivery Ratio ($PDR$) and the frame detection probability by the jammer with real world LoRa transceivers.
The authors from \cite{aras},\cite{aras2} highlighted that the long Time On Air ($TOA$) of LoRa gives a bigger opportunity window for the jammer, especially with high modulation orders.
Physical layer mitigation techniques were proposed to reduce jamming effectiveness such as frequency hopping scheme that was presented in \cite{ahmar}.
To the best of authors knowledge, traditional jamming (mainly Band Jamming ($BJ$) and Tone Jamming ($TJ$)) impact on LoRa has not been investigated yet.
Although smart jammers (\textit{e.g.} malicious LoRa user) are more efficient, traditional jammers are still a threat for LoRa networks.
\textcolor{black}{Indeed, traditional jammers use usually low-cost devices and require minimal setup procedures.
They can be then easily implemented and need therefore to be tackled.}
This paper focuses on the LoRa victim node. 
We propose to investigate the effect of both $BJ$ and $TJ$ on LoRa signals, the performance associated and a simple jamming detection scheme leveraging LoRa physical layer characteristics is derived.
This gives the ability to a LoRa node to alert the presence of a jammer in the close environment.
The main contributions of this paper are:

\begin{itemize}
    \item An analysis of both $BJ$ and $TJ$ applied on LoRa signals that reveals the good LoRa's resiliency against these jamming attacks.
    \item A simple and efficient $BJ$ and $TJ$ detection scheme enabling more security in LoRa networks.
\end{itemize}

The remainder of the paper is organized as follows.
In Section \ref{sec:LoRaOverview}, a brief review of LoRa physical layer is performed.
Section \ref{sec:JamModels} presents $BJ$ and $TJ$ models and the different jamming strategies are discussed.
Section \ref{sec:JamEffect} evaluates the impact of $BJ$ \& $TJ$ on LoRa signals while Section \ref{sec:JamDetect} introduces the jammer detection scheme.
Simulation results are presented in Section \ref{sec:Simu} to assess jamming impact on Symbol Error Rate ($SER$) performance and  jamming detection capability.
Finally, Section \ref{sec:conclusion} concludes the paper.

\section{LoRa modulation overview}\label{sec:LoRaOverview}

\subsection{LoRa waveforms}\label{subsec:LoRaSignals}

In the literature, LoRa waveforms are of the type of Chirp Spread Spectrum ($CSS$) signals.
These signals rely on sine waves with Instantaneous Frequency ($IF$) that varies linearly with time over frequency range $f \in [-B/2,B/2]$ ($B\in\{125,250,500\}$~kHz) and time range $t \in [0,T]$ ($T$ denotes the symbol period).
This basic signal is called an \emph{up-chirp} or \emph{down-chirp} when frequency respectively increases or decreases over time.
A LoRa symbol consists of $SF$ bits ($SF \in \{7,8,\ldots,12\}$) leading to an $M$-ary modulation with $M = 2^{SF} \in \{128,256,\ldots,4096\}$.
\textcolor{black}{The symbol duration $T$ is defined as $T = M/B$.
This gives $T \approx \{1,2,\ldots,32\}$ ms for $B = 125$ kHz and $SF = \{7,\ldots,12\}$.
We may see that LoRa has quite long symbol duration compared to other modulation schemes such as $OFDM$ where symbols usually last only tens of $\mu s$.}
In the discrete-time signal model, the Nyquist sampling rate ($F_s=1/T_s$) is used (\textit{i.e.} $T_s  = 1/B = T/M$) to reduce complexity.
The signal symbol has then $M$ samples.
Each symbol $a \in \{0,1,\ldots,M-1\}$ is mapped to an \emph{up-chirp} that is temporally shifted by $\tau_a = a T_s$ periods.
We may notice that a temporal shift $\tau_a$ conducts to shift by $aB/M=a/(MT_s)=a/T$ the $IF$.
The modulo operation is applied to ensure that $IF$ remains in the interval $[-B/2,B/2]$.
This behavior is the heart of $CSS$ process.
A mathematical expression of LoRa waveform sampled at $t = k T_s$ has been derived in\cite{chiani} :

\begin{equation}
    x(kT_s;a) \triangleq x_a[k] = e^{2j\pi k \left( \frac{a}{M} - \frac{1}{2} + \frac{k}{2M} \right)} \quad k = 0,1,\ldots,M-1
    \label{eqxa}
\end{equation}

We may see that an \emph{up-chirp} is actually a LoRa waveform with symbol index $a = 0$, written $x_0[k]$.
Its conjugate $x_0^*[k]$ is then a \emph{down-chirp}. 

\subsection{LoRa demodulation scheme}\label{LoRaDemodScheme}

Reference \cite{vangelista} proposed a simple and efficient solution to demodulate LoRa signals.
In Additive White Gaussian Noise ($AWGN$) channel, the demodulation process is based on the Maximum Likelihood ($ML$) detection scheme.
The received signal is:
\begin{equation}
    r[k] = x_a[k] + w[k]
\end{equation}
with $w[k]$ a complex $AWGN$ with zero-mean and variance $\sigma^2 = E[|w[k]|^2]$.
The Signal to Noise Ratio ($SNR$) is defined as $SNR = 1/\sigma^2$.
$ML$ detector aims to select index $\widehat{a}$ that maximizes the scalar product $\langle r[k],x_n[k] \rangle$ for $n = 0,1,\ldots,M-1$ defined as:

\begin{equation}
\begin{split}
    \langle r[k],x_n[k] \rangle &= \sum_{k=0}^{M-1} r[k] x_n^*[k]\\
    &= \sum_{k=0}^{M-1} \underbrace{(x_a[k] + w[k]) x_0^*[k]}_{\Tilde{r}[k]} e^{-j2\pi\frac{n}{M}k} \\
    &= \Tilde{R}[n] = \Tilde{X}_a[n] + \Tilde{W}[n] = M \delta[n-a] + \Tilde{W}[n]
\end{split}
\end{equation}

with $\Tilde{X}_a[k] = DFT\{x_a[k] x_0^*[k]\} = M \delta[n-a]$ and $\Tilde{W}[k] = DFT\{w[k] x_0^*[k]\} \sim \mathcal{CN}(0,\sigma_w^2)$, $\sigma_w^2 = M \sigma^2$.
\textcolor{black}{$DFT\{.\}$ denotes the Discrete Fourier Transform ($DFT$) function.}
The demodulation stage proceeds with two simple operations:

\begin{itemize}
    \item multiply the received signal by the \emph{down-chirp} $x_0^*[k]$, also called dechirping,
    \item compute $\Tilde{R}[n]$, the $DFT$ of $\Tilde{r}[k]$ and select the discrete frequency index $\widehat{a}$ that maximizes  $\Tilde{R}[n]$.
    \label{eqDemodLoRa}
\end{itemize}

This way, the dechirp process merges all the signal energy in a unique frequency bin $a$ and can be easily retrieved by taking the magnitude or square magnitude (non-coherent detection) of $\Tilde{R}[n]$.
The symbol detection is then:

\begin{equation}
    \begin{split}
        \widehat{a}_{NCOH} &= \underset{n}{\argmax} \quad |\Tilde{R}[n]|^2 \equiv \underset{n}{\argmax} \quad |\Tilde{R}[n]|
    \end{split}
    \label{eqDemodLoRaNCOH}
\end{equation}

\section{Traditional Jammer models}\label{sec:JamModels}

This section briefly reviews traditional jammer models and the main strategies that can be adopted by the jammer.

\subsection{Band Jamming}

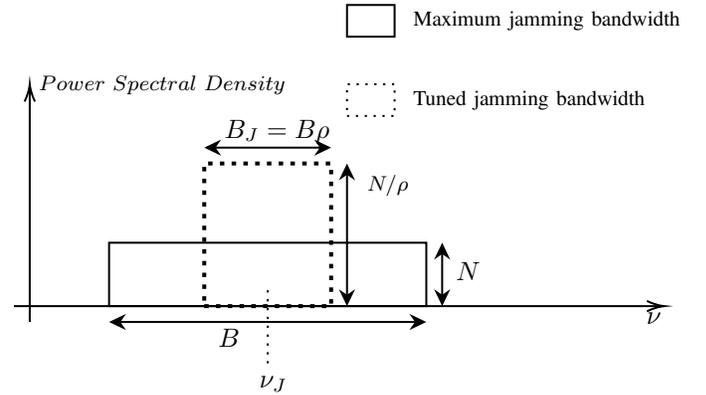
\begin{figure}[ht]
  \centering
  \input{figBJ}
  \caption{Band Jamming model illustration.}
\label{fig:BJ}
\end{figure}

$BJ$ model is presented in Figure \ref{fig:BJ}.
It adds a jamming signal $w_J[k]$, usually an $AWGN$, in the data signal bandwidth.
The jamming signal bandwidth may be restricted to a fraction of the data signal bandwidth as:

\begin{equation}
    B_J = B \times \rho, \quad \rho \in ]0;1]
\end{equation}

When $\rho = 1$, the jammer covers the entire useful bandwidth and is called Full Band Jamming ($FBJ$).
For $\rho \neq 1$, the jammer is a Partial Band Jammer ($PBJ$).
% \textcolor{blue}{The jamming power is fixed and constant that is reducing $B_J$ increases band magnitude as:}
\textcolor{black}{The total jamming power is fixed to $\sigma_J^2 = B \times N$ where $N$ is the spectral density level and $B$ the total available bandwidth, as depicted in Figure \ref{fig:BJ}.
$\rho$ is the fraction of $B$ to be covered by the jammer.
As total available jamming power is fixed, reducing $\rho$ increases therefore the spectral density level by a factor $1/\rho$.
}

% \textcolor{blue}{
% \begin{equation}
%     \sigma_J^2 = \frac{N}{\rho} \times B_J = B \times N
% \end{equation}
% }

The jammer can also tune the bandwidth position \textit{i.e.} center frequency $\nu_J$.

\subsection{Tone Jamming}

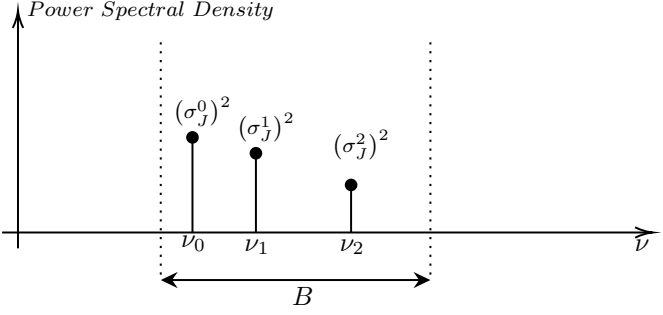
\begin{figure}[ht]
  \centering
  \input{figTJ}
  \caption{Tone Jamming model illustration for $V=3$.}
\label{fig:TJ}
\end{figure}

$TJ$ adds several sine waveforms in the useful bandwidth as depicted in Figure \ref{fig:TJ}.
The expression of the discrete jamming signal sampled at $t=k T_s$ is then:

\begin{equation}
\begin{split}
    & s_{TJ}[k] = \sum_{v=0}^{V-1} s_{TJ}^v[k] = \sum_{v=0}^{V-1} \sigma_J^v e^{2j\pi \nu_v k + j\phi_v} \\
    & \nu_v = \frac{u_v}{M}, \quad u_v \in ]0;M-1]
\end{split}
    \label{eqMTJ}
\end{equation}

With $\phi_v$ the initial phase of the $v$th tone signal and uniformly distributed over $[0;2\pi]$.
When $V=1$ the jamming signal is Single Tone Jamming ($STJ$) and Multi Tone Jamming ($MTJ$) otherwise.
From \eqref{eqMTJ} the jammer can tune each sine waveform power.
An optimal power strategy for the jammer is to adopt an uniform scheme \cite{li} \textit{i.e.} $(\sigma_J^v)^2 = \sigma_J^2/V$.
This scheme is considered in this paper.

\subsection{Jamming behaviors}

In the literature, the jammer has mainly three different behaviors: the constant, reactive and random jammers.
The constant jammer has the maximum impact on the victim but is not energy efficient and easily detectable.
This behavior is therefore usually ignored at the expense of the reactive jammer that is passive most of the time and sends a jamming signal only when detecting the target signal to be jammed.
This behavior is a good energy/impact trade-off and is very likely to be found in real threats.
We consider this strategy in the rest of the paper.
Random jamming is less effective than reactive jamming but difficult to detect due to its random nature.
Depending on the jammer behavior, the received signal at the victim LoRa node follows the next four hypotheses:

\begin{itemize}
    \item $H_0$ : $r[k] = w[k]$
    \item $H_1$ : $r[k] = x_a[k] + w[k] + s[k]$
    \item $H_2$ : $r[k] = x_a[k] + w[k]$
    \item $H_3$ : $r[k] = w[k] + s[k]$
\end{itemize}

Hypotheses $H_1$ and $H_3$ are only valid for the constant jammer.
The reactive jammer enables hypotheses $H_0$ and $H_1$ while random jammer enables the four hypotheses.

\section{$BJ$ \& $TJ$ effect on LoRa signals}\label{sec:JamEffect}

The effect of jamming is only present for $H_1$.
We denote the Noise Jamming Ratio ($NJR$) as $NJR = \sigma^2/\sigma_J^2$.
The received signal at the victim LoRa node after dechirp process and $DFT$ is:

\begin{equation}
    \Tilde{R}[n] = \Tilde{X}_a[n] + \Tilde{W}[n] + \Tilde{S}[n]
    \label{eqSigRecH1}
\end{equation}

The term $\Tilde{S}[n]$ in \eqref{eqSigRecH1} introduces interference.
We focus on this term for the study of jamming.
The term $\Tilde{S}[n]$ is renamed as $\tilde{S}_{BJ}[n]$ for $BJ$ and $\Tilde{S}_{TJ}[n]$ for $TJ$.

\subsection{$BJ$ effect}\label{subsec:BJEffect}

\begin{figure}[ht]
  \centering
  \input{figBJEffect}
  \caption{Partial Band Jamming effect on LoRa $DFT$.}
\label{fig:BJEffect}
\end{figure}
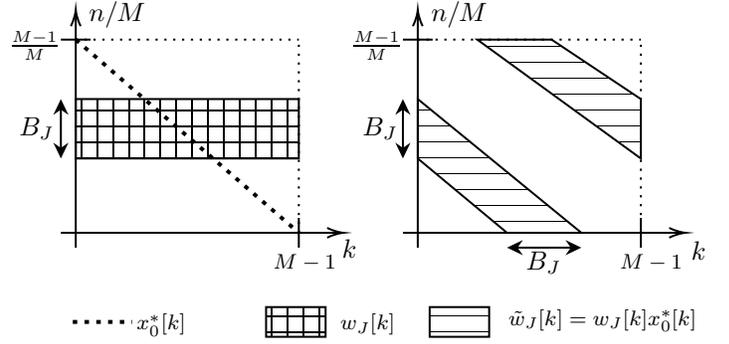

The effect is illustrated in Figure \ref{fig:BJEffect}.
We may see that $PBJ$ has the same effect as $FBJ$ thanks to dechirp operation.
% With $FBJ$, the noise power is spread uniformly over the entire frequency/time plane with power level denoted as $\sigma_{FBJ}^2$.
% Each frequency component has then the same energy $E_{FBJ} = \sigma_{FBJ}^2 \times (M-1)$.
% $PBJ$ increases the noise power by $1/\rho$ factor in the restricted band \textit{i.e.} $\sigma_{PBJ}^2 = \sigma_{FBJ}^2/\rho$.
% The energy of each frequency is therefore $E_{PBJ} = \sigma_{PBJ}^2 \rho (M-1) = \sigma_{FBJ}^2 \times (M-1) = E_{FBJ}$.
The $DFT$ output will have then an equivalent noise power of $\sigma_{BJ}^2 = \sigma_w^2 + M\sigma_J^2$ with limited impact on performance as it will be highlighted in Section \ref{sec:Simu}.

\subsection{$TJ$ effect}\label{subsec:TJEffect}

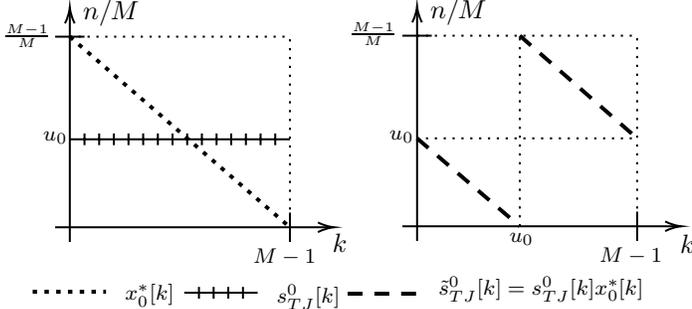
\begin{figure}[ht]
  \centering
  \input{figTJEffect}
  \caption{Single Tone Jamming effect on LoRa $DFT$.}
\label{fig:TJEffect}
\end{figure}

$TJ$ effect is illustrated similarly as $BJ$ in Figure \ref{fig:TJEffect}.
We consider for instance integer $u_v$ values.
The dechirp and $DFT$ operations lead to an effect equivalent as computing the $DFT$ of a LoRa symbol with value $a=u_v$ and modulated by a \emph{down-chirp}.
The interference term is then:

\begin{equation}
    \Tilde{S}_{TJ}[n] = \sum_{v=0}^{V-1} DFT\{x_{u_v}^*[k]\}
    \label{eqMTJDFT}
\end{equation}

Without loss of generality we consider $V=1$ to evaluate \eqref{eqMTJDFT}.
Developing \eqref{eqMTJDFT} yields:

\begin{equation}
    \Tilde{S}_{STJ}[n] = \sqrt{\sigma_J^2} \sum_{k=0}^{M-1} e^{2j\pi \frac{-k^2 + (M+2u_0-2n)k}{2M}}
    \label{eqMTJDFT1}
\end{equation}

The sum in \eqref{eqMTJDFT1} is in the form of a Generalized Quadratic Gaussian Sum ($GQGS$) resolution task.
The $GQGS$ is defined as \cite{berndt}:

\begin{equation}
    G(\eta,\epsilon,\gamma) = \sum_{x=0}^{|\gamma|-1} e\left(\frac{\eta x^2 + \epsilon x}{\gamma}\right), \quad e(x) = e^{2j\pi x}
    \label{eqGQGS}
\end{equation}

% With $\eta$, $\epsilon$ and $\gamma$ natural numbers.
With $\eta$, $\epsilon$ and $\gamma$ integers.
When $\eta$ odd, $\epsilon$ even and $\gamma$ a power of 2, $G(\eta,\epsilon,\gamma)$ is:

\begin{equation}
    G(\eta,\epsilon,\gamma) = e\left(-\frac{\frac{{\epsilon}^2}{4\eta}}{\gamma}\right) G(\eta,|\gamma|)
    \label{eqQGS}
\end{equation}

with $G(\eta,|\gamma|) = (1 + j^{\eta}) \sqrt{|\gamma|}$.
By identification between \eqref{eqMTJDFT1} and \eqref{eqGQGS}, $\eta=-1$, $\epsilon=M+2u_0-2n$ and $\gamma=2M$ for LoRa.
$\eta$ is odd, $\epsilon$ is even and $\gamma$ a power of 2 so \eqref{eqQGS} is valid.
$G(\eta,|\gamma|)$ is then for LoRa $G(-1,2M) = (1 - j) \sqrt{2M}$.
We note that the sum in \eqref{eqMTJDFT1} has a range of $k = 0,\ldots,M-1$, contrarily to \eqref{eqGQGS} that supposes to have a sum from $x=0$ to $x=2M-1$.
It can be normalized knowing that $\sum_{k=0}^{2M-1} e^{2j\pi \frac{\eta k^2 + \epsilon k}{\gamma}} = 2 \sum_{k=0}^{M-1} e^{2j\pi \frac{\eta k^2 + \epsilon k}{\gamma}}$ with $e^{2j\pi \frac{\eta k^2 + \epsilon k}{\gamma}} = e^{2j\pi \frac{\eta (k+M)^2 + \epsilon (k+M)}{\gamma}}$, $k = 0,\ldots,M-1$.
$\Tilde{S}_{STJ}[n]$ is then:

\begin{equation}
    \Tilde{S}_{STJ}[n] = \sqrt{\sigma_J^2} \frac{G(-1,\epsilon_0,2M)}{2}, \quad \epsilon_0 = M+2u_0-2n
    \label{eqSTJDFTGQGS}
\end{equation}

$\Tilde{S}_{TJ}[n]$ is finally for any $V$:

\begin{equation}
    \Tilde{S}_{TJ}[n] = \sqrt{\sigma_J^2/V} \sum_{v=0}^{V-1} \frac{G(-1,\epsilon_v,2M)}{2}, \quad \epsilon_v = M+2u_v-2n
    \label{eqMTJDFTGQGS}
\end{equation}

When $u_v$ is not integer, the computation of \eqref{eqMTJDFT1} is not possible with \eqref{eqMTJDFTGQGS}.
% \textcolor{blue}{The sum must be then computed manually.}
\textcolor{black}{The sum must be then computed numerically for each $n$ value.}

\begin{figure}[ht]
  \centering
  \includegraphics[width=0.49\textwidth]{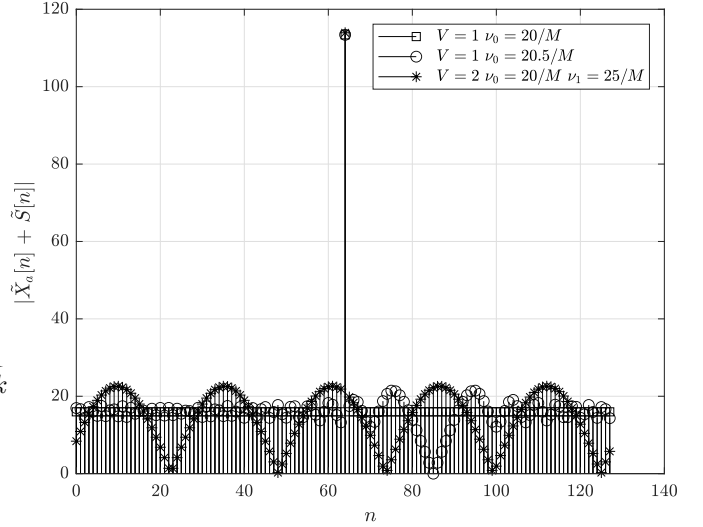}
  \caption{$DFT$ output of received LoRa plus $TJ$ signal without noise for $V=\{1,2\}$ and different sine waveform of frequencies $u_v$.
  $SF=7$, $\sigma_J^2 = 2$ and $a=M/2$.}
\label{fig:STJDFT}
\end{figure}

Figure \ref{fig:STJDFT} presents the $DFT$ output of received LoRa signal contaminated by $TJ$, for $SF=7$.
This $SF$ value is considered for the rest of the paper.
The noise term $\Tilde{W}[n]$ is neglected for convenience.
% The noise term $\Tilde{W}[n]$ is neglected for convenience and the $DFT$ square magnitude detector is considered in this figure to bring to light $TJ$ behavior.
When $V=1$ and $u_0$ integer, the $DFT$ magnitude is equal for each bin at $n \neq a$ and of value $\sqrt{M \sigma_J^2} = 16$ in this example ($\sigma_J^2=2$).
When $u_0$ is not integer, the $DFT$ experiences deformations.
These deformations are maximum when $u_0 = \left \lfloor{u_0}\right \rfloor + 0.5$ and has an oscillation behavior centered around $\sqrt{M \sigma_J^2}$, as showed in the figure.
The oscillation is null at $n = M/2 + \left \lfloor{u_0}\right \rfloor + 1$.
When $V = 2$, a sine modulation wise behavior appears with frequency of approximately $|u_0-u_1| \mod M$ and shifted circularly by $\min(u_0,u_1)$ positions.
The $DFT$ output is quite unpredictable for $V>1$ and $u_v$ not integer.
We may see that the bin magnitude at $n=a$ depends on $a$.
That is, certain $a$ values depending on $u_v$ will reduce or increase the expected magnitude of $M$ without jamming.
More precisely, we may see the following condition leading to performance improvement or degradation:

\begin{equation}
\text{if}
    \begin{cases}
    \Re \{ {\Tilde{S}[a]}\} > 0, & |\Tilde{R}[a]| > M : \text{performance improvement} \\
    \Re \{ {\Tilde{S}[a]}\} < 0, & |\Tilde{R}[a]| < M : \text{performance degradation}
\end{cases}
\label{eq:}
\end{equation}

The symbols minimizing and maximizing performance are denoted $a_{min}$ and $a_{max}$ and are derived as:

\begin{equation}
\begin{split}
    & a_{min} = \underset{n}{\argmin} \quad {\Re \{ {\Tilde{S}[n]}\}} \\
    & a_{max} = \underset{n}{\argmax} \quad {\Re \{ {\Tilde{S}[n]}\}}
\end{split}
\end{equation}

\begin{figure}[ht]
  \centering
  \includegraphics[width=0.49\textwidth]{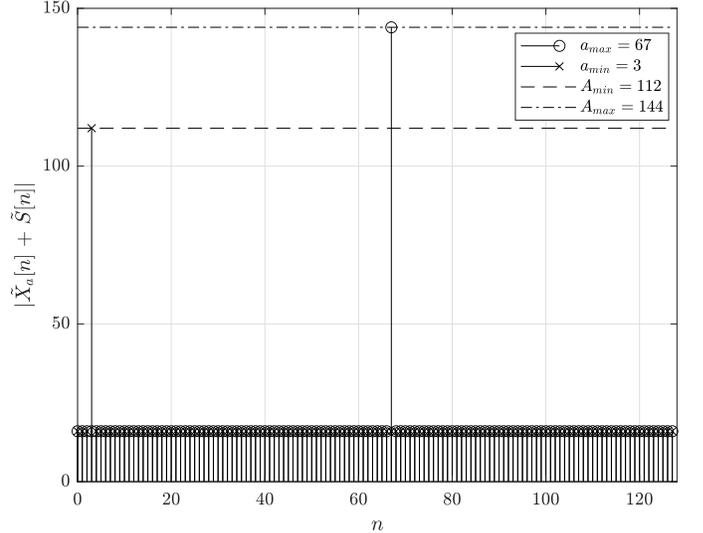}
  \caption{$DFT$ output of received LoRa plus $STJ$ signal without noise for $V=1$, $u_0=20$, $\sigma_J^2 = 2$ and $SF=7$.}
\label{fig:STJDFTSymbOptMin}
\end{figure}

An example of $DFT$ output with symbols $a=a_{max}$ and $a=a_{min}$ is showed in Figure \ref{fig:STJDFTSymbOptMin}, with $u_0=20$ and $\sigma_J^2 = 2$.
In this case, $a_{max}=67$ and $a_{min}=3$.
Moreover, the performance gain and loss are respectively $\Gamma^+ = \Tilde{R}[a_{max}] - M = \sqrt{M \sigma_J^2}$ and $\Gamma^- = M - \Tilde{R}[a_{min}] = \sqrt{M \sigma_J^2}$, for $V=1$.
When $V>1$, the performance gain/loss does not hold exactly the same behavior with $\Gamma^+ > \Gamma^-$ or $\Gamma^+ < \Gamma^-$, depending on $u_v$ values.

\section{Jammer detection scheme}\label{sec:JamDetect}

\subsection{LoRa $DFT$ $PDF$s}

The jamming detection scheme is performed in the frequency domain \textit{i.e.} the LoRa dechirped $DFT$ to keep simple implementation.
We may recall the Probability Density Functions ($PDF$) of $|\Tilde{R}[n]|$ for the two hypotheses $H_0$ and $H_1$ (reactive jammer), noted as $|\Tilde{R}_{H_0}[n]|$, $|\Tilde{R}^{BJ}_{H_1}[n]|$ and $|\Tilde{R}^{TJ}_{H_1}[n]|$ for $BJ$ and $TJ$, respectively.
In $H_0$, only $AWGN$ is present.
Its statistic is not changed by the dechirp process.
We can easily conclude that Random Variable ($RV$) $X_{H_0}$ of $|\Tilde{R}_{H_0}[n]|$ follows a Rayleigh distribution $f_{X_{H_0}}(t) = Rayl(t,b_{X_{H_0}})$ with $b_{X_{H_0}} = \sqrt{M \sigma^2/2}$ the scale parameter.

The term $\Tilde{S}[n]$ in $H_1$ leads to a Rayleigh $PDF$ $f_{X^{BJ}_{H_1}}(t) = Rayl(t,b_{X^{BJ}_{H_1}})$ with $b_{X^{BJ}_{H_1}} = \sqrt{M (\sigma^2+\sigma_J^2)/2}$, for $BJ$ and $n \neq a$.
$TJ$ throws a Rician $PDF$ to $|\Tilde{R}^{TJ}_{H_1}[n]| \sim X^{TJ}_{H_1}$, $f_{X^{TJ}_{H_1}}(t) = Rice(t,\mu_{X_{H_1}^{TJ}},\sigma_{X^{TJ}_{H_1}})$ with non centrality parameter $\mu_{X_{H_1}^{TJ}} = \sqrt{M \sigma_J^2}$ and scale parameter $\sigma_{X^{TJ}_{H_1}} = b_{X_{H_0}}$, $n \neq a$.

\subsection{Jammer detector}

A simple solution to detect the $BJ$ or $TJ$ jammer is to compute the following normalized quantity test:

\begin{equation}
    z = \sum_{l=0}^{L-1} \frac{|\Tilde{R}[n_l]|}{b_{X_{H_0}}}, \quad n_l \neq a
    \label{eq:z}
\end{equation}

As the jammer is reactive, the LoRa node can estimate regularly $w[k]$ variance during silence periods \textit{i.e.} in $H_0$ hypothesis and leverage this information to detect the jammer.
From \eqref{eq:z}, the receiver chooses randomly $L$ frequency indexes different from $n=a$.
Indeed, the $PDF$ of $|\Tilde{R}_{H_1}[a]|$ depends on $a$ for both $BJ$ and $TJ$, an information not available.
To mitigate this situation, the receiver can eliminate the $N_{\lambda}$ frequency bins that are above a certain threshold.
The threshold is designed with Neyman-Pearson ($NP$) criterion.
The False Alarm Probability ($FAP$) $P_{fa}$ is fixed and the threshold is derived by:

\begin{equation}
    \lambda = F^{-1}_{Rayl}(1-P_{fa};\sqrt{M \sigma^2/2})
    \label{eq:lambda}
\end{equation}

with $F_{Rayl}^{-1}(.;.)$ denoting the inverse Cumulative Density Function ($CDF$) of Rayleigh $RV$.
We note $z_{H_0}$ and $z_{H_1}$ the quantity test in $H_0$ and $H_1$ hypotheses, respectively.
It is worth-noting that $z_{H_0}$ is a sum of Rayleigh $RV$s.
The evaluation of $PDF$ and $CDF$ of a sum of Rayleigh $RV$s has been studied in the literature.
The authors from \cite{schwartz} derived a closed-form approximation based on Small Argument Approximation ($SAA$) approach.
Their solution is widely used but has the drawback to introduce bias as $L$ grows.
More recent studies proposed a closed-form expression but limited to small $L$ values, $L \in \{2,\ldots,16\}$ in \cite{hu} for example.
To have more flexibility, we choose $SAA$ technique.
From \cite{schwartz}, $PDF$ $Z_{H_0}$ is:

\begin{equation}
\begin{split}
    f_{Z_{H_0}}^{SAA}(t) &= \sum_{l=0}^{L-1} X_{H_0}^l = \frac{t^{2L-1} e^{-\frac{t^2}{2b_{Z_{H_0}^{SAA}}}}}{2^{L-1} (b_{Z_{H_0}^{SAA}})^{L} (L-1)!^1} \\
    b_{Z_{H_0}^{SAA}} &= \frac{1}{L}[(2L-1)!^2]^{1/L} \\
    % (2L-1)!! &= (2L-1)(2L-3) \ldots 3 \times 1
\end{split}
\label{eq:zH0SAA}
\end{equation}

\textcolor{black}{where
\begin{equation}
    x!^c = 1 \times (1+c) \times (1+2c) \times \ldots \times x
\end{equation}
}

Its $CDF$ is:

\begin{equation}
    F_{Z_{H_0}}^{SAA}(t) = 1 - e^{-\frac{t^2}{2b_{Z_{H_0}^{SAA}}}} \sum_{l=0}^{L-1} \frac{\left( \frac{t^2}{2b_{Z_{H_0}^{SAA}}} \right)^l}{l!^1}
    \label{eq:zH0SAACDF}
\end{equation}

In $H_1$ and $BJ$, $z_{H_1}^{BJ}$ follows the same $PDF$ as $z_{H_0}$ but with different parameter $b_{Z_{H_1}^{BJ,SAA}} = \frac{1 + \frac{\sigma_J^2}{\sigma^2}}{L}[(2L-1)!^2]^{1/L}$ as the noise $DFT$ has variance $\sigma_{BJ}^2$.
The associated $CDF$ is $F_{Z_{H_1}}^{BJ,SAA}$.
In $TJ$ case, $z_{H_1}^{TJ}$ is a sum of Rician $RV$s.
Similar research to evaluate sum of Rician $RV$s has been performed \cite{hu2},\cite{salcedo} but are still limited to small $L$ values (up to $L=10$ in \cite{salcedo}) and reduced number of possible $NJR$ values, $NJR_{dB} \in \{-7,-5,-3,-1\}$ in \cite{hu2}.
This limits the application to our LoRa jammer detector.
We decide to use instead the approximation of the Rician distribution when $V=1$.
If $NJR_{dB} \rightarrow \infty$, $f_{Z_{H_1}}^{STJ}$ approaches $f_{Z_{H_0}}^{SAA}$ and noted $f_{Z_{H_1}}^{STJ^+}$.
If $NJR_{dB} \rightarrow -\infty$, $f_{Z_{H_1}}^{STJ}$ is a normal distribution noted $f_{Z_{H_1}}^{STJ^-}$:

\begin{equation}
\begin{split}
    & f_{Z_{H_1}}^{STJ^-}(t) = \mathcal{N}(t,\mu_{Z_{H_1}^{STJ^-}},\sigma_{Z_{H_1}^{STJ^-}}) \\
    & \mu_{Z_{H_1}^{STJ^-}} = \frac{\sqrt{M (\sigma^2/2 + \sigma_J^2)L}}{b_{X_{H_0}}}, \quad \sigma_{Z_{H_1}^{STJ^-}} = 1
\end{split}
\label{eq:zH1TJPDF}
\end{equation}

The $CDF$ are $F_{Z_{H_1}^{STJ^-}}$ and $F_{Z_{H_1}^{STJ^+}}$.
Intermediate $NJR_{dB}$ values will lead to reasonably small bias.
When $V>1$, the $PDF$ has not analytical expression and must be therefore numerically computed.
$CDF$ is noted $F_{Z_{H_1}}^{MTJ}$.
Finally, the receiver detects the jammer with a threshold identically designed as $\lambda$ with fixed $FAP$:

\begin{equation}
    \lambda_{SAA} = \left(F_{Z_{H_0}}^{SAA}\right)^{-1}(1-P_{fa}^{SAA})
    \label{eq:lambdaSAA}
\end{equation}

\begin{equation}
    z \quad\mathop{\gtreqless}_{H_0}^{H_1} \quad \lambda_{SAA}
    \label{eq:Detect}
\end{equation}

\begin{figure}[ht]
  \centering
  \includegraphics[width=0.49\textwidth]{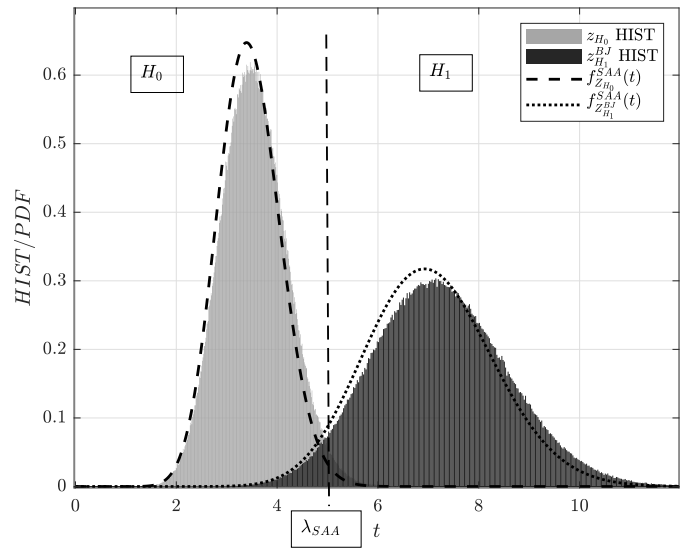}
  \caption{Theoretical $PDF$s and histograms of quantities test $z_{H_0}$ and $z_{H_1}^{BJ}$ for $BJ$.
  $NJR_{dB} = -5$, $SNR_{dB} = -5$, $L=8$ and $\rho = 0.6$.
  Histograms have indexes $t_h$ normalized such that $t = t_h/\sqrt{L}$ as stated in \cite{hu}.}
\label{fig:PDF_HIST_BJ_NJRm5_SNRm5_L8_rho06}
\end{figure}

An illustration for $BJ$ case of theoretical $PDF$s and histograms for both $H_0$ and $H_1$ hypotheses are depicted in Figure \ref{fig:PDF_HIST_BJ_NJRm5_SNRm5_L8_rho06}.

\section{simulation results}\label{sec:Simu}

This section provides Monte-Carlo simulation results to evaluate $BJ$ and $TJ$ performance impact on $SER$ and assess the jammer detection capability.
% The simulations are performed with $SF=7$.
To simulate $BJ$, an $AWGN$ is generated and filtered according to $\rho$ and $\sigma_J^2$ constraints.

\subsection{$PBJ$ and $MTJ$ performance impact on $SER$}

The simulations are performed with respect to Signal Jamming Ratio ($SJR$) $SJR = 1/\sigma_J^2$.
The $SNR$ is fixed to $SNR_{dB}=-8$.

\subsubsection{$PBJ$ performance impact on $SER$}

\begin{figure}[ht]
  \centering
  \includegraphics[width=0.49\textwidth]{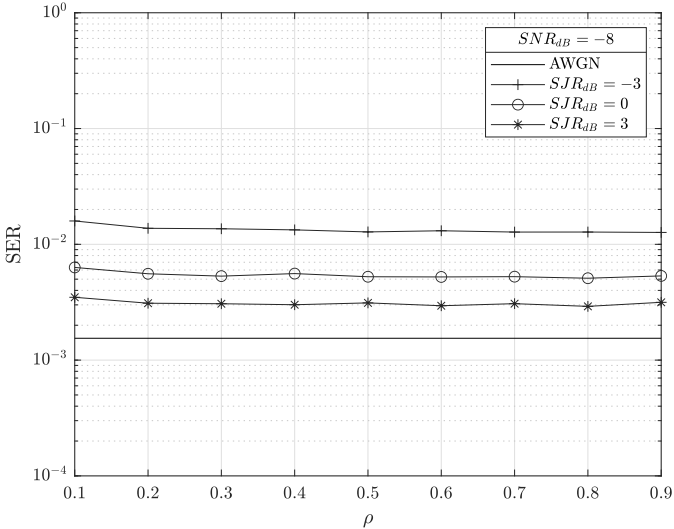}
  \caption{$PBJ$ $SER$ performance depending on $\rho \in \{0.1,0.2,\ldots,0.9\}$
  $SNR_{dB}=-8$ and $SJR_{dB} \in \{-3,0,3\}$.}
\label{fig:perfs_BJ_rho_var_SNRm8_NJRm3_0_3}
\end{figure}

Figure \ref{fig:perfs_BJ_rho_var_SNRm8_NJRm3_0_3} highlights the impact of $\rho$ on performance as mentioned in Section \ref{subsec:BJEffect}.
$SJR_{dB} \in \{-3,0,3\}$ and $AWGN$ performance showed as comparison.
Higher $SJR_{dB}$ values slowly reduce performance with a $SER$ difference of roughly $1.1 \times 10^{-2}$ between $SJR_{dB}=-3$ and $SJR_{dB}=3$.
% \textcolor{blue}{The more $SJR_{dB}$ the closer to $AWGN$ performance are.}
\textcolor{black}{We may see in the Figure that $\rho$ has virtually no impact on performance whatever $SJR$ is.
This confirms performance predictions made in Section \ref{subsec:BJEffect}, as $\sigma_J^2$ is fixed.}
It is obvious that $BJ$ is not a good strategy for the jammer.

\subsubsection{$MTJ$ performance impact on $SER$}
\begin{figure}[ht]
  \centering
  \includegraphics[width=0.49\textwidth]{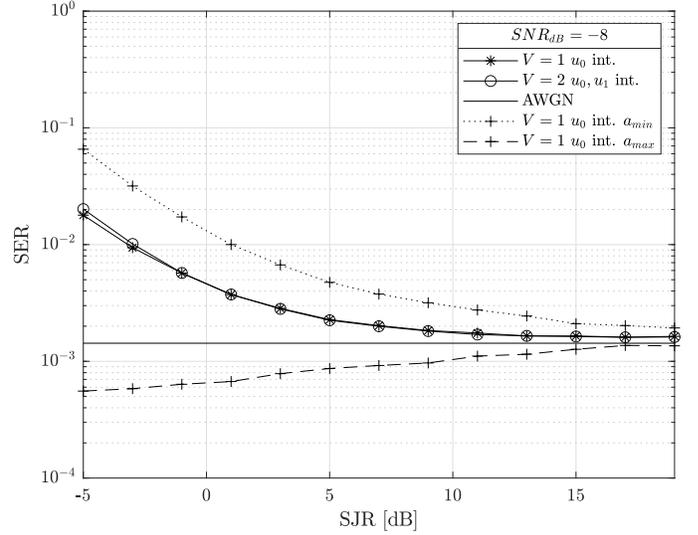}
  \caption{$MTJ$ $SER$ performance as a function of $SJR$.
  $SNR_{dB}=-8$, $u_v$ integer and considering $a_{min}$ and $a_{max}$ symbol values.}
\label{fig:perfs_TJ_SJR_var_SNRm8_J_1_2_amin_amax}
\end{figure}

For $MTJ$ performance impact evaluation, $u_v$ is integer and chosen uniformly in $u_v \in \{0,\ldots,M-1\}$ at each Monte-Carlo trial.
Figure \ref{fig:perfs_TJ_SJR_var_SNRm8_J_1_2_amin_amax} points out two interesting performance results.
First, $V$ has no particular influence on performance when $u_v$ is integer.
This also true for $u_v$ non integer.
Indeed, the periodic $DFT$ output behavior for $u_v$ non integer as depicted in Figure \ref{fig:STJDFT} has average magnitude values around $\sqrt{M \sigma_J^2}$, the value when $u_v$ is integer.
Statistically, for $a$ random and uniform over $[0;M-1]$, this does not influence performance.
Second, $a$ value has a huge impact on performance.
The symbol minimizing performance $a_{min}$ leads to very poor performance that reaches $AWGN$ one only from $SJR_{dB}=20$.
Interestingly, $a_{max}$ does not improve so much performance, with only a gain of about $1.1 \times 10^{-3}$ at $SJR_{dB}=-5$.

\subsection{Jammer detection performance}

We compare in this section jammer detection performance for several $NJR_{dB}$ and $L$ values.
Theoretical $CDF$s $F_{Z_{H_0}}^{SAA}$ and $F_{Z_{H_1}}^{BJ,SAA}$ for $BJ$ are computed using Equation \eqref{eq:zH0SAACDF}.
Depending on $V$, $F_{Z_{H_1}^{STJ^-}}$, $F_{Z_{H_1}^{STJ^+}}$ or $F_{Z_{H_1}^{MTJ}}$ are computed for $TJ$ case.
Note that $F_{Z_{H_1}^{MTJ}}$ is numerically computed based on its $PDF$ with Monte-Carlo trials.
$\lambda_{SAA}$ is also computed numerically as \cite{hu} does not provide inverse $CDF$.
The Miss Detection Probability ($MDP$) for $BJ$ and $TJ$ are respectively computed as:

\begin{equation}
    P_{md}^{BJ} = \int_{-\infty}^{\lambda_{SAA}} f_{Z_{H_1}}^{BJ,SAA}(t) dt = F_{Z_{H_1}}^{BJ,SAA}(\lambda_{SAA})
    \label{eq:PmdBJ}
\end{equation}

\begin{equation}
    P_{md}^{STJ} = \int_{-\infty}^{\lambda_{SAA}} f_{Z_{H_1}}^{STJ^{+/-}}(t) dt = F_{Z_{H_1}}^{STJ^{+/-}}(\lambda_{SAA})
    \label{eq:PmdSTJ}
\end{equation}

\begin{equation}
    P_{md}^{MTJ} = \int_{-\infty}^{\lambda_{SAA}} f_{Z_{H_1}}^{MTJ}(t) dt = F_{Z_{H_1}}^{MTJ}(\lambda_{SAA})
    \label{eq:PmdMTJ}
\end{equation}

\subsubsection{Theoretical performance}

\begin{figure}[ht]
  \centering
  \includegraphics[width=0.49\textwidth]{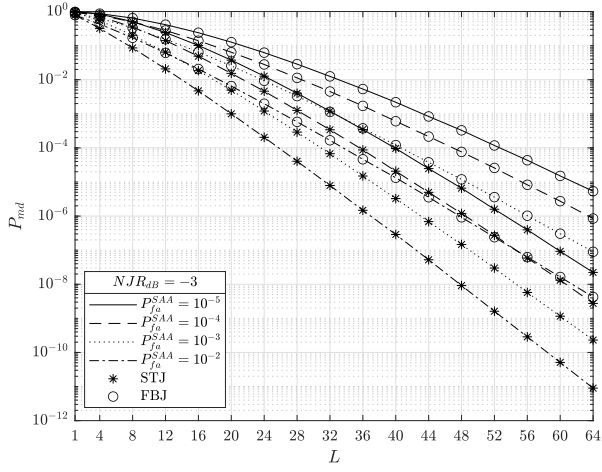}
  \caption{Theoretical $FBJ$ and $STJ$ $P_{md}$ as a function of $L$ for different $P_{fa}^{SAA} \in \{10^{-2},10^{-3},10^{-4},10^{-5}\}$.
  $NJR_{dB}=-3$.}
\label{fig:perfs_Pmd_BJ_STJ_L_vect_NJRm3}
\end{figure}

\begin{figure}[ht]
  \centering
  \includegraphics[width=0.49\textwidth]{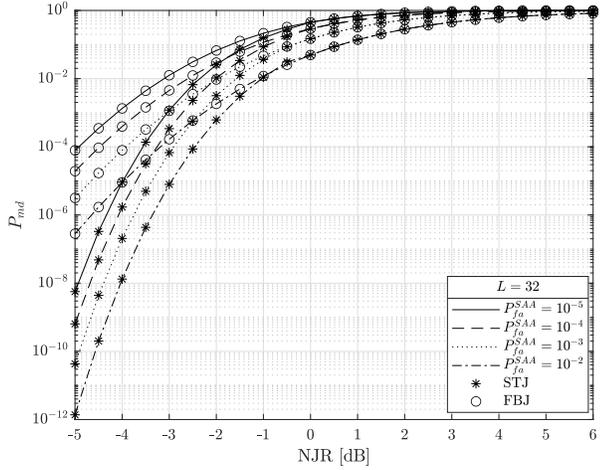}
  \caption{Theoretical $FBJ$ and $STJ$ $P_{md}$ as a function of $NJR$ for different $P_{fa}^{SAA} \in \{10^{-2},10^{-3},10^{-4},10^{-5}\}$.
  $L=32$.}
\label{fig:perfs_Pmd_BJ_STJ_NJR_vect_L32}
\end{figure}

Theoretical $P_{md}$ performance as a function of $L$ and $NJR$ are plotted in Figures \ref{fig:perfs_Pmd_BJ_STJ_L_vect_NJRm3} and \ref{fig:perfs_Pmd_BJ_STJ_NJR_vect_L32}, respectively, for $FBJ$ and $STJ$.
$\rho$ then equals 1.
We may see that as $L$ increases or $NJR$ decreases, the $P_{md}$ reduces.
Consequently, the jammer detection performance increases.
It can be easily explained by the fact that for $L=1$, taking a single frequency bin is insufficient to make a proper decision.
Ideally, $L = M-N_{\lambda}$ (the bin magnitude at $n=a$ is ignored).
However, the theoretical/simulation bias will be more important for large $L$ due to $SAA$ limitation and therefore the obtained $P_{md}$ will be slightly different, as it will be highlighted in next section.
If $NJR$ is to high, the jammer power is overlooked in the noise floor leading to impossible detection.
A trade-off between $P_{fa}^{SAA}$ and $P_{md}$ is necessary as less false alarm will increase non detection.
It can also be seen that $STJ$ detection outperforms $FBJ$ detection.
A $P_{md}$ factor of about 390 at $L=64$ and $P_{fa}^{SAA}=10^{-3}$ is experienced in Figure \ref{fig:perfs_Pmd_BJ_STJ_L_vect_NJRm3}.

\subsubsection{Theoretical versus Simulation performance}

\begin{figure}[ht]
  \centering
  \includegraphics[width=0.49\textwidth]{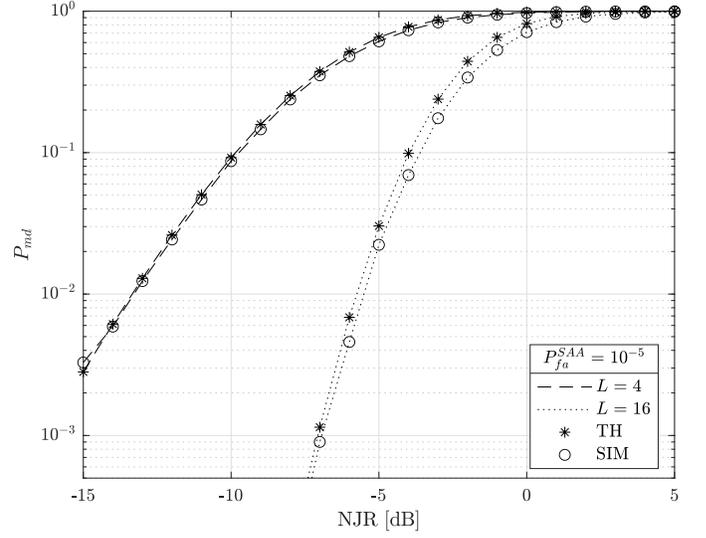}
  \caption{Theoretical versus simulation $P_{md}$ for $PBJ$ as a function of $NJR$ for different $P_{fa}^{SAA} = 10^{-5}$, $L \in \{4,16\}$, $SNR_{dB}=0$ and $\rho = 0.6$.}
\label{fig:perfs_PBJ_Pmd_NJR_Pfavect_th_sim_L416}
\end{figure}

We finally compare simulation against theoretical $P_{md}$.
$PBJ$ simulation are performed with $\rho = 0.6$.
$MTJ$ is evaluated for $V \in \{1,3\}$ with $\nu_v$ values $\nu_0=0.711$, $\nu_1=0.812$ and $\nu_2=0.273$.
From Figure \ref{fig:perfs_PBJ_Pmd_NJR_Pfavect_th_sim_L416}, we may see that $L=16$ introduces a slightly higher bias compared with $L=4$, as expected, in favor of higher $P_{md}$ performance.

\begin{figure}[ht]
  \centering
  \includegraphics[width=0.49\textwidth]{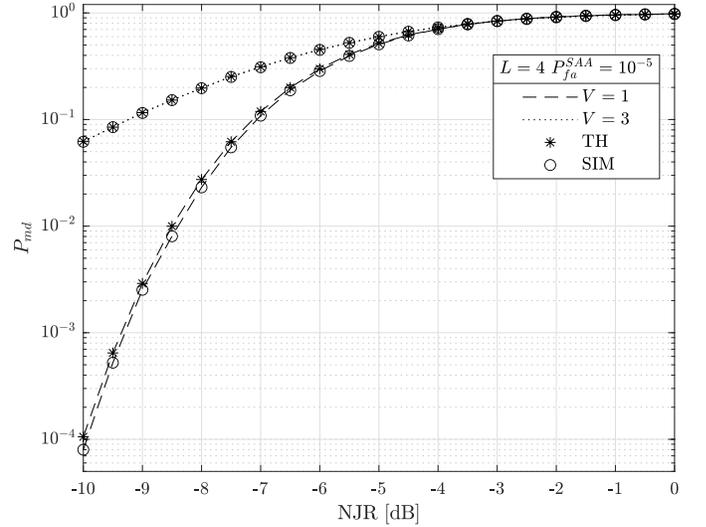}
  \caption{Theoretical versus simulation $P_{md}$ for $STJ$ ($V=1$) and $MTJ$ ($V=3$) as a function of $NJR$ for different $P_{fa}^{SAA} = 10^{-5}$, $SNR_{dB}=0$ and $L=4$.}
\label{fig:perfs_STJ_Pmd_NJR_Pfavect_th_sim_J_13_L4}
\end{figure}

From Figure \ref{fig:perfs_STJ_Pmd_NJR_Pfavect_th_sim_J_13_L4}, we remark that the theory/simulation bias is reduced for $V=3$.
Indeed, computing numerically $F_{Z_{H_1}^{MTJ}}$ removes CDF/histogram bias.
% Computing numerically $F_{Z_{H_1}^{MTJ}}$ remove indeed $CDF$/histogram bias.
$MTJ$ detection experiences a performance hit with $V=3$ but is still more efficient than $PBJ$.
At $NJR_{dB}=-10$ and $P_{fa}^{SAA}=10^{-5}$, $BJ$ has a $P_{md}$ of roughly $9 \times 10^{-2}$ against $6 \times 10^{-2}$ for $MTJ$.
Higher $V$ values does not change $P_{md}$ performance \textit{i.e.} $P_{md}^{v-1} \approx P_{md}^{v}$ for $V > 1$.

\section{Conclusion}\label{sec:conclusion}

In this paper, analysis of both $BJ$ and $TJ$ on LoRa signals was carried out.
We pointed out that $PBJ$ has virtually same effect as $FBJ$.
It is therefore equivalent to an additional source of $AWGN$ leading then to small $SER$ performance degradation.
We also highlighted that $TJ$ parameters $V$ and $u_v$ has negligible impact on $SER$ performance.
We can conclude that LoRa is quite robust to $BJ$ and $TJ$.
We also developed a simple scheme to detect efficiently $BJ$ and $TJ$.
The method leverages traditional basic LoRa processing without adding a burden on complexity.
Overall, $TJ$ detector performs better than $BJ$, even when considering $MTJ$ scenario even though $MTJ$ detection appears to be less efficient than $TJ$.
\textcolor{black}{This work can be further extended.
For example, real tests on LoRa transceivers can be performed to assess the article conclusions drawn beforehand in more realistic conditions.
Furthermore, the theoretical impact of a multi-path channel on these jamming schemes can be explored.
Temporal and frequency de-synchronizations impact may be also investigated.}
% \textcolor{blue}{This work may be extended to a more realistic scenario, multi-path channels for example.}

\section*{Acknowledgment}

This work was jointly supported by the Brest Institute of Computer Science and Mathematics ($IBNM$) CyberIoT Chair of Excellence of the University of Brest, the Brittany Region and the “Pôle d’Excellence Cyber”.

\bibliographystyle{IEEEtran}
\bibliography{IEEEabrv,biblio}

\end{document}

%% file: figBJ.tex
\tikzset{every picture/.style={line width=0.75pt}} %set default line width to 0.75pt        

\begin{tikzpicture}[x=0.75pt,y=0.75pt,yscale=-0.8,xscale=0.8]

%Straight Lines [id:da26640128346304626] 
\draw    (50,200) -- (458,200) ;
\draw [shift={(460,200)}, rotate = 180] [color={rgb, 255:red, 0; green, 0; blue, 0 }  ][line width=0.75]    (10.93,-3.29) .. controls (6.95,-1.4) and (3.31,-0.3) .. (0,0) .. controls (3.31,0.3) and (6.95,1.4) .. (10.93,3.29)   ;
%Shape: Rectangle [id:dp37367800772246285] 
\draw   (110,160) -- (310,160) -- (310,200) -- (110,200) -- cycle ;
%Shape: Rectangle [id:dp3565114582162032] 
\draw  [color={rgb, 255:red, 0; green, 0; blue, 0 }  ,draw opacity=1 ][dash pattern={on 1.69pt off 2.76pt}][line width=1.5]  (170,110) -- (250,110) -- (250,200) -- (170,200) -- cycle ;
%Straight Lines [id:da27106555494328366] 
\draw    (113,210) -- (307,210) ;
\draw [shift={(310,210)}, rotate = 180] [fill={rgb, 255:red, 0; green, 0; blue, 0 }  ][line width=0.08]  [draw opacity=0] (10.72,-5.15) -- (0,0) -- (10.72,5.15) -- (7.12,0) -- cycle    ;
\draw [shift={(110,210)}, rotate = 0] [fill={rgb, 255:red, 0; green, 0; blue, 0 }  ][line width=0.08]  [draw opacity=0] (10.72,-5.15) -- (0,0) -- (10.72,5.15) -- (7.12,0) -- cycle    ;
%Straight Lines [id:da9080315158936612] 
\draw    (260,197) -- (260,113) ;
\draw [shift={(260,110)}, rotate = 450] [fill={rgb, 255:red, 0; green, 0; blue, 0 }  ][line width=0.08]  [draw opacity=0] (10.72,-5.15) -- (0,0) -- (10.72,5.15) -- (7.12,0) -- cycle    ;
\draw [shift={(260,200)}, rotate = 270] [fill={rgb, 255:red, 0; green, 0; blue, 0 }  ][line width=0.08]  [draw opacity=0] (10.72,-5.15) -- (0,0) -- (10.72,5.15) -- (7.12,0) -- cycle    ;
%Straight Lines [id:da34681363915211616] 
\draw    (320,197) -- (320,163) ;
\draw [shift={(320,160)}, rotate = 450] [fill={rgb, 255:red, 0; green, 0; blue, 0 }  ][line width=0.08]  [draw opacity=0] (10.72,-5.15) -- (0,0) -- (10.72,5.15) -- (7.12,0) -- cycle    ;
\draw [shift={(320,200)}, rotate = 270] [fill={rgb, 255:red, 0; green, 0; blue, 0 }  ][line width=0.08]  [draw opacity=0] (10.72,-5.15) -- (0,0) -- (10.72,5.15) -- (7.12,0) -- cycle    ;
%Straight Lines [id:da5724857585217789] 
\draw    (173,100) -- (247,100) ;
\draw [shift={(250,100)}, rotate = 180] [fill={rgb, 255:red, 0; green, 0; blue, 0 }  ][line width=0.08]  [draw opacity=0] (10.72,-5.15) -- (0,0) -- (10.72,5.15) -- (7.12,0) -- cycle    ;
\draw [shift={(170,100)}, rotate = 0] [fill={rgb, 255:red, 0; green, 0; blue, 0 }  ][line width=0.08]  [draw opacity=0] (10.72,-5.15) -- (0,0) -- (10.72,5.15) -- (7.12,0) -- cycle    ;
%Straight Lines [id:da015109753533634196] 
\draw  [dash pattern={on 0.84pt off 2.51pt}]  (210,190) -- (210,240) ;
%Straight Lines [id:da5994989807930837] 
\draw    (60,210) -- (60,62) ;
\draw [shift={(60,60)}, rotate = 450] [color={rgb, 255:red, 0; green, 0; blue, 0 }  ][line width=0.75]    (10.93,-3.29) .. controls (6.95,-1.4) and (3.31,-0.3) .. (0,0) .. controls (3.31,0.3) and (6.95,1.4) .. (10.93,3.29)   ;
%Shape: Rectangle [id:dp03700968695862472] 
\draw   (260,10) -- (290,10) -- (290,30) -- (260,30) -- cycle ;
%Shape: Rectangle [id:dp05637099283662206] 
\draw  [dash pattern={on 0.84pt off 2.51pt}] (260,60) -- (290,60) -- (290,80) -- (260,80) -- cycle ;

% Text Node
\draw (271,115.4) node [anchor=north west][inner sep=0.75pt]  [font=\footnotesize]  {$N /\rho $};
% Text Node
\draw (177,212.4) node [anchor=north west][inner sep=0.75pt]    {$B$};
% Text Node
\draw (181,80.4) node [anchor=north west][inner sep=0.75pt]    {$B_{J} =B\rho $};
% Text Node
\draw (327,170.4) node [anchor=north west][inner sep=0.75pt]    {$N$};
% Text Node
\draw (203,242.4) node [anchor=north west][inner sep=0.75pt]    {$\nu _{J}$};
% Text Node
\draw (447,202.4) node [anchor=north west][inner sep=0.75pt]    {$\nu $};
% Text Node
\draw (64,52.4) node [anchor=north west][inner sep=0.75pt]  [font=\footnotesize]  {$Power\ Spectral\ Density$};
% Text Node
\draw (300,12) node [anchor=north west][inner sep=0.75pt]  [font=\footnotesize] [align=left] {Maximum jamming bandwidth};
% Text Node
\draw (300,62) node [anchor=north west][inner sep=0.75pt]  [font=\footnotesize] [align=left] {Tuned jamming bandwidth};
\end{tikzpicture}

%% file: figTJ.tex
\tikzset{every picture/.style={line width=0.75pt}} %set default line width to 0.75pt        

\begin{tikzpicture}[x=0.75pt,y=0.75pt,yscale=-0.8,xscale=0.8]

%Straight Lines [id:da35146437880233106] 
\draw    (50,200) -- (458,200) ;
\draw [shift={(460,200)}, rotate = 180] [color={rgb, 255:red, 0; green, 0; blue, 0 }  ][line width=0.75]    (10.93,-3.29) .. controls (6.95,-1.4) and (3.31,-0.3) .. (0,0) .. controls (3.31,0.3) and (6.95,1.4) .. (10.93,3.29)   ;
%Straight Lines [id:da4745505841118458] 
\draw    (170,140) -- (170,200) ;
\draw [shift={(170,140)}, rotate = 90] [color={rgb, 255:red, 0; green, 0; blue, 0 }  ][fill={rgb, 255:red, 0; green, 0; blue, 0 }  ][line width=0.75]      (0, 0) circle [x radius= 3.35, y radius= 3.35]   ;
%Straight Lines [id:da6084565043888119] 
\draw    (210,150) -- (210,200) ;
\draw [shift={(210,150)}, rotate = 90] [color={rgb, 255:red, 0; green, 0; blue, 0 }  ][fill={rgb, 255:red, 0; green, 0; blue, 0 }  ][line width=0.75]      (0, 0) circle [x radius= 3.35, y radius= 3.35]   ;
%Straight Lines [id:da9329912583006874] 
\draw    (270,170) -- (270,200) ;
\draw [shift={(270,170)}, rotate = 90] [color={rgb, 255:red, 0; green, 0; blue, 0 }  ][fill={rgb, 255:red, 0; green, 0; blue, 0 }  ][line width=0.75]      (0, 0) circle [x radius= 3.35, y radius= 3.35]   ;
%Straight Lines [id:da42751304241916044] 
\draw    (60,210) -- (60,62) ;
\draw [shift={(60,60)}, rotate = 450] [color={rgb, 255:red, 0; green, 0; blue, 0 }  ][line width=0.75]    (10.93,-3.29) .. controls (6.95,-1.4) and (3.31,-0.3) .. (0,0) .. controls (3.31,0.3) and (6.95,1.4) .. (10.93,3.29)   ;
%Straight Lines [id:da4392419546485984] 
\draw    (153,230) -- (294.41,230) -- (317,230) ;
\draw [shift={(320,230)}, rotate = 180] [fill={rgb, 255:red, 0; green, 0; blue, 0 }  ][line width=0.08]  [draw opacity=0] (10.72,-5.15) -- (0,0) -- (10.72,5.15) -- (7.12,0) -- cycle    ;
\draw [shift={(150,230)}, rotate = 0] [fill={rgb, 255:red, 0; green, 0; blue, 0 }  ][line width=0.08]  [draw opacity=0] (10.72,-5.15) -- (0,0) -- (10.72,5.15) -- (7.12,0) -- cycle    ;
%Straight Lines [id:da9537078218443991] 
\draw  [dash pattern={on 0.84pt off 2.51pt}]  (150,80) -- (150,230) ;
%Straight Lines [id:da26931728941541455] 
\draw  [dash pattern={on 0.84pt off 2.51pt}]  (320,80) -- (320,230) ;

% Text Node
\draw (447,202.4) node [anchor=north west][inner sep=0.75pt]    {$\nu $};
% Text Node
\draw (161,200.4) node [anchor=north west][inner sep=0.75pt]    {$\nu _{0}$};
% Text Node
\draw (201,202.4) node [anchor=north west][inner sep=0.75pt]    {$\nu _{1}$};
% Text Node
\draw (261,202.4) node [anchor=north west][inner sep=0.75pt]    {$\nu _{2}$};
% Text Node
\draw (156,112.4) node [anchor=north west][inner sep=0.75pt]  [font=\footnotesize]  {$\left( \sigma _{J}^{0}\right)^{2}$};
% Text Node
\draw (196,122.4) node [anchor=north west][inner sep=0.75pt]  [font=\footnotesize]  {$\left( \sigma _{J}^{1}\right)^{2}$};
% Text Node
\draw (256,133.4) node [anchor=north west][inner sep=0.75pt]  [font=\footnotesize]  {$\left( \sigma _{J}^{2}\right)^{2}$};
% Text Node
\draw (64,52.4) node [anchor=north west][inner sep=0.75pt]  [font=\footnotesize]  {$Power\ Spectral\ Density$};
% Text Node
\draw (231,232.4) node [anchor=north west][inner sep=0.75pt]    {$B$};

\end{tikzpicture}

%% file: figBJEffect.tex
\tikzset{
pattern size/.store in=\mcSize, 
pattern size = 5pt,
pattern thickness/.store in=\mcThickness, 
pattern thickness = 0.3pt,
pattern radius/.store in=\mcRadius, 
pattern radius = 1pt}\makeatletter
\pgfutil@ifundefined{pgf@pattern@name@_q6cgq915p}{
\pgfdeclarepatternformonly[\mcThickness,\mcSize]{_q6cgq915p}
{\pgfqpoint{-\mcThickness}{-\mcThickness}}
{\pgfpoint{\mcSize}{\mcSize}}
{\pgfpoint{\mcSize}{\mcSize}}
{\pgfsetcolor{\tikz@pattern@color}
\pgfsetlinewidth{\mcThickness}
\pgfpathmoveto{\pgfpointorigin}
\pgfpathlineto{\pgfpoint{\mcSize}{0}}
\pgfpathmoveto{\pgfpointorigin}
\pgfpathlineto{\pgfpoint{0}{\mcSize}}
\pgfusepath{stroke}}}
\makeatother

% Pattern Info
 
\tikzset{
pattern size/.store in=\mcSize, 
pattern size = 5pt,
pattern thickness/.store in=\mcThickness, 
pattern thickness = 0.3pt,
pattern radius/.store in=\mcRadius, 
pattern radius = 1pt}
\makeatletter
\pgfutil@ifundefined{pgf@pattern@name@_i6plpk55k lines}{
\pgfdeclarepatternformonly[\mcThickness,\mcSize]{_i6plpk55k}
{\pgfqpoint{0pt}{0pt}}
{\pgfpoint{\mcSize+\mcThickness}{\mcSize+\mcThickness}}
{\pgfpoint{\mcSize}{\mcSize}}
{\pgfsetcolor{\tikz@pattern@color}
\pgfsetlinewidth{\mcThickness}
\pgfpathmoveto{\pgfpointorigin}
\pgfpathlineto{\pgfpoint{\mcSize}{0}}
\pgfusepath{stroke}}}
\makeatother

% Pattern Info
 
\tikzset{
pattern size/.store in=\mcSize, 
pattern size = 5pt,
pattern thickness/.store in=\mcThickness, 
pattern thickness = 0.3pt,
pattern radius/.store in=\mcRadius, 
pattern radius = 1pt}
\makeatletter
\pgfutil@ifundefined{pgf@pattern@name@_0uas4zh7n lines}{
\pgfdeclarepatternformonly[\mcThickness,\mcSize]{_0uas4zh7n}
{\pgfqpoint{0pt}{0pt}}
{\pgfpoint{\mcSize+\mcThickness}{\mcSize+\mcThickness}}
{\pgfpoint{\mcSize}{\mcSize}}
{\pgfsetcolor{\tikz@pattern@color}
\pgfsetlinewidth{\mcThickness}
\pgfpathmoveto{\pgfpointorigin}
\pgfpathlineto{\pgfpoint{\mcSize}{0}}
\pgfusepath{stroke}}}
\makeatother

% Pattern Info
 
\tikzset{
pattern size/.store in=\mcSize, 
pattern size = 5pt,
pattern thickness/.store in=\mcThickness, 
pattern thickness = 0.3pt,
pattern radius/.store in=\mcRadius, 
pattern radius = 1pt}\makeatletter
\pgfutil@ifundefined{pgf@pattern@name@_r839halmn}{
\pgfdeclarepatternformonly[\mcThickness,\mcSize]{_r839halmn}
{\pgfqpoint{-\mcThickness}{-\mcThickness}}
{\pgfpoint{\mcSize}{\mcSize}}
{\pgfpoint{\mcSize}{\mcSize}}
{\pgfsetcolor{\tikz@pattern@color}
\pgfsetlinewidth{\mcThickness}
\pgfpathmoveto{\pgfpointorigin}
\pgfpathlineto{\pgfpoint{\mcSize}{0}}
\pgfpathmoveto{\pgfpointorigin}
\pgfpathlineto{\pgfpoint{0}{\mcSize}}
\pgfusepath{stroke}}}
\makeatother

% Pattern Info
 
\tikzset{
pattern size/.store in=\mcSize, 
pattern size = 5pt,
pattern thickness/.store in=\mcThickness, 
pattern thickness = 0.3pt,
pattern radius/.store in=\mcRadius, 
pattern radius = 1pt}
\makeatletter
\pgfutil@ifundefined{pgf@pattern@name@_zvforw6d4 lines}{
\pgfdeclarepatternformonly[\mcThickness,\mcSize]{_zvforw6d4}
{\pgfqpoint{0pt}{0pt}}
{\pgfpoint{\mcSize+\mcThickness}{\mcSize+\mcThickness}}
{\pgfpoint{\mcSize}{\mcSize}}
{\pgfsetcolor{\tikz@pattern@color}
\pgfsetlinewidth{\mcThickness}
\pgfpathmoveto{\pgfpointorigin}
\pgfpathlineto{\pgfpoint{\mcSize}{0}}
\pgfusepath{stroke}}}
\makeatother
\tikzset{every picture/.style={line width=0.75pt}} %set default line width to 0.75pt        

\begin{tikzpicture}[x=0.75pt,y=0.75pt,yscale=-0.75,xscale=0.75]
%uncomment if require: \path (0,256); %set diagram left start at 0, and has height of 256

%Straight Lines [id:da5729796869090948] 
\draw    (22,160) -- (210,160) ;
\draw [shift={(212,160)}, rotate = 180] [color={rgb, 255:red, 0; green, 0; blue, 0 }  ][line width=0.75]    (10.93,-3.29) .. controls (6.95,-1.4) and (3.31,-0.3) .. (0,0) .. controls (3.31,0.3) and (6.95,1.4) .. (10.93,3.29)   ;
%Straight Lines [id:da9376806406462921] 
\draw    (32,170) -- (32,12) ;
\draw [shift={(32,10)}, rotate = 450] [color={rgb, 255:red, 0; green, 0; blue, 0 }  ][line width=0.75]    (10.93,-3.29) .. controls (6.95,-1.4) and (3.31,-0.3) .. (0,0) .. controls (3.31,0.3) and (6.95,1.4) .. (10.93,3.29)   ;
%Straight Lines [id:da49834991861814637] 
\draw    (22,30) -- (42,30) ;
%Straight Lines [id:da41825984630007795] 
\draw    (182,170) -- (182,150) ;
%Straight Lines [id:da5046483975318559] 
\draw [line width=1.5]  [dash pattern={on 1.69pt off 2.76pt}]  (32,30) -- (182,160) ;
%Straight Lines [id:da2674313248116613] 
\draw    (252,160) -- (440,160) ;
\draw [shift={(442,160)}, rotate = 180] [color={rgb, 255:red, 0; green, 0; blue, 0 }  ][line width=0.75]    (10.93,-3.29) .. controls (6.95,-1.4) and (3.31,-0.3) .. (0,0) .. controls (3.31,0.3) and (6.95,1.4) .. (10.93,3.29)   ;
%Straight Lines [id:da5420989426943779] 
\draw    (262,170) -- (262,12) ;
\draw [shift={(262,10)}, rotate = 450] [color={rgb, 255:red, 0; green, 0; blue, 0 }  ][line width=0.75]    (10.93,-3.29) .. controls (6.95,-1.4) and (3.31,-0.3) .. (0,0) .. controls (3.31,0.3) and (6.95,1.4) .. (10.93,3.29)   ;
%Straight Lines [id:da43076343879234114] 
\draw    (412,170) -- (412,150) ;
%Straight Lines [id:da5736487462903239] 
\draw    (252,30) -- (272,30) ;
%Straight Lines [id:da9250364668343778] 
\draw  [dash pattern={on 0.84pt off 2.51pt}]  (412,160) -- (412,30) ;
%Straight Lines [id:da03479067629964261] 
\draw  [dash pattern={on 0.84pt off 2.51pt}]  (182,160) -- (182,30) ;
%Straight Lines [id:da06624662355672095] 
\draw  [dash pattern={on 0.84pt off 2.51pt}]  (32,30) -- (182,30) ;
%Straight Lines [id:da7890010857490135] 
\draw  [dash pattern={on 0.84pt off 2.51pt}]  (262,30) -- (412,30) ;
%Shape: Rectangle [id:dp288223095127059] 
\draw  [pattern=_q6cgq915p,pattern size=6pt,pattern thickness=0.75pt,pattern radius=0pt, pattern color={rgb, 255:red, 0; green, 0; blue, 0}] (32,70) -- (182,70) -- (182,110) -- (32,110) -- cycle ;
%Straight Lines [id:da03795591584613267] 
\draw    (22,73) -- (22,107) ;
\draw [shift={(22,110)}, rotate = 270] [fill={rgb, 255:red, 0; green, 0; blue, 0 }  ][line width=0.08]  [draw opacity=0] (10.72,-5.15) -- (0,0) -- (10.72,5.15) -- (7.12,0) -- cycle    ;
\draw [shift={(22,70)}, rotate = 90] [fill={rgb, 255:red, 0; green, 0; blue, 0 }  ][line width=0.08]  [draw opacity=0] (10.72,-5.15) -- (0,0) -- (10.72,5.15) -- (7.12,0) -- cycle    ;
%Straight Lines [id:da3047278018511832] 
\draw    (252,73) -- (252,107) ;
\draw [shift={(252,110)}, rotate = 270] [fill={rgb, 255:red, 0; green, 0; blue, 0 }  ][line width=0.08]  [draw opacity=0] (10.72,-5.15) -- (0,0) -- (10.72,5.15) -- (7.12,0) -- cycle    ;
\draw [shift={(252,70)}, rotate = 90] [fill={rgb, 255:red, 0; green, 0; blue, 0 }  ][line width=0.08]  [draw opacity=0] (10.72,-5.15) -- (0,0) -- (10.72,5.15) -- (7.12,0) -- cycle    ;
%Straight Lines [id:da977950507233277] 
\draw    (325,170) -- (369,170) ;
\draw [shift={(372,170)}, rotate = 180] [fill={rgb, 255:red, 0; green, 0; blue, 0 }  ][line width=0.08]  [draw opacity=0] (10.72,-5.15) -- (0,0) -- (10.72,5.15) -- (7.12,0) -- cycle    ;
\draw [shift={(322,170)}, rotate = 0] [fill={rgb, 255:red, 0; green, 0; blue, 0 }  ][line width=0.08]  [draw opacity=0] (10.72,-5.15) -- (0,0) -- (10.72,5.15) -- (7.12,0) -- cycle    ;
%Shape: Polygon [id:ds8656120296037759] 
\draw  [pattern=_i6plpk55k,pattern size=6pt,pattern thickness=0.75pt,pattern radius=0pt, pattern color={rgb, 255:red, 0; green, 0; blue, 0}] (352,30) -- (412,70) -- (412,110) -- (302,30) -- (302,30) -- cycle ;
%Shape: Polygon [id:ds9675205023990805] 
\draw  [pattern=_0uas4zh7n,pattern size=6pt,pattern thickness=0.75pt,pattern radius=0pt, pattern color={rgb, 255:red, 0; green, 0; blue, 0}] (372,160) -- (322,160) -- (262,110) -- (262,70) -- (262,70) -- cycle ;
%Shape: Rectangle [id:dp6423558523015132] 
\draw  [pattern=_r839halmn,pattern size=6pt,pattern thickness=0.75pt,pattern radius=0pt, pattern color={rgb, 255:red, 0; green, 0; blue, 0}] (160,210) -- (200,210) -- (200,230) -- (160,230) -- cycle ;
%Straight Lines [id:da23532445366503452] 
\draw [line width=1.5]  [dash pattern={on 1.69pt off 2.76pt}]  (30,220) -- (70,220) ;
%Shape: Rectangle [id:dp1564371909673743] 
\draw  [pattern=_zvforw6d4,pattern size=6pt,pattern thickness=0.75pt,pattern radius=0pt, pattern color={rgb, 255:red, 0; green, 0; blue, 0}] (270,210) -- (310,210) -- (310,230) -- (270,230) -- cycle ;

% Text Node
\draw (39,2.4) node [anchor=north west][inner sep=0.75pt]    {$n/M$};
% Text Node
\draw (209,162.4) node [anchor=north west][inner sep=0.75pt]    {$k$};
% Text Node
\draw (163,172.4) node [anchor=north west][inner sep=0.75pt]  [font=\footnotesize]  {$M-1$};
% Text Node
\draw (-15,22.4) node [anchor=north west][inner sep=0.75pt]  [font=\footnotesize]  {$\frac{M-1}{M}$};
% Text Node
\draw (396,172.4) node [anchor=north west][inner sep=0.75pt]  [font=\footnotesize]  {$M-1$};
% Text Node
\draw (215,22.4) node [anchor=north west][inner sep=0.75pt]  [font=\footnotesize]  {$\frac{M-1}{M}$};
% Text Node
\draw (444,163.4) node [anchor=north west][inner sep=0.75pt]    {$k$};
% Text Node
\draw (269,2.4) node [anchor=north west][inner sep=0.75pt]    {$n/M$};
% Text Node
\draw (71,212.4) node [anchor=north west][inner sep=0.75pt]  [font=\footnotesize]  {$x^{*}_{0}[ k]$};
% Text Node
\draw (321,209.4) node [anchor=north west][inner sep=0.75pt]  [font=\footnotesize]  {$\tilde{w}_{J}[ k] =w_{J}[ k] x^{*}_{0}[ k]$};
% Text Node
\draw (-8,80.4) node [anchor=north west][inner sep=0.75pt]    {$B_{J}$};
% Text Node
\draw (223,80.4) node [anchor=north west][inner sep=0.75pt]    {$B_{J}$};
% Text Node
\draw (333,170.4) node [anchor=north west][inner sep=0.75pt]    {$B_{J}$};
% Text Node
\draw (208,212.4) node [anchor=north west][inner sep=0.75pt]  [font=\footnotesize]  {$w_{J}[ k]$};
\end{tikzpicture}

%% file: figTJEffect.tex
\tikzset{every picture/.style={line width=0.75pt}} %set default line width to 0.75pt        

\begin{tikzpicture}[x=0.75pt,y=0.75pt,yscale=-0.74,xscale=0.74]
%uncomment if require: \path (0,248); %set diagram left start at 0, and has height of 248

%Straight Lines [id:da002723344300225472] 
\draw    (24,165) -- (212,165) ;
\draw [shift={(214,165)}, rotate = 180] [color={rgb, 255:red, 0; green, 0; blue, 0 }  ][line width=0.75]    (10.93,-3.29) .. controls (6.95,-1.4) and (3.31,-0.3) .. (0,0) .. controls (3.31,0.3) and (6.95,1.4) .. (10.93,3.29)   ;
%Straight Lines [id:da2132300500347899] 
\draw    (34,175) -- (34,17) ;
\draw [shift={(34,15)}, rotate = 450] [color={rgb, 255:red, 0; green, 0; blue, 0 }  ][line width=0.75]    (10.93,-3.29) .. controls (6.95,-1.4) and (3.31,-0.3) .. (0,0) .. controls (3.31,0.3) and (6.95,1.4) .. (10.93,3.29)   ;
%Straight Lines [id:da11289462137393569] 
\draw    (24,35) -- (44,35) ;
%Straight Lines [id:da7101568326979555] 
\draw    (184,175) -- (184,155) ;
%Straight Lines [id:da694201386741246] 
\draw [line width=1.5]  [dash pattern={on 1.69pt off 2.76pt}]  (34,35) -- (184,165) ;
%Straight Lines [id:da42197761506553855] 
\draw [color={rgb, 255:red, 0; green, 0; blue, 0 }  ,draw opacity=1 ]   (34,105) -- (184,105) (44,101) -- (44,109)(54,101) -- (54,109)(64,101) -- (64,109)(74,101) -- (74,109)(84,101) -- (84,109)(94,101) -- (94,109)(104,101) -- (104,109)(114,101) -- (114,109)(124,101) -- (124,109)(134,101) -- (134,109)(144,101) -- (144,109)(154,101) -- (154,109)(164,101) -- (164,109)(174,101) -- (174,109) ;
%Straight Lines [id:da10650739675138476] 
\draw    (260.67,164.33) -- (448.67,164.33) ;
\draw [shift={(450.67,164.33)}, rotate = 180] [color={rgb, 255:red, 0; green, 0; blue, 0 }  ][line width=0.75]    (10.93,-3.29) .. controls (6.95,-1.4) and (3.31,-0.3) .. (0,0) .. controls (3.31,0.3) and (6.95,1.4) .. (10.93,3.29)   ;
%Straight Lines [id:da35587467683644336] 
\draw    (270.67,174.33) -- (270.67,16.33) ;
\draw [shift={(270.67,14.33)}, rotate = 450] [color={rgb, 255:red, 0; green, 0; blue, 0 }  ][line width=0.75]    (10.93,-3.29) .. controls (6.95,-1.4) and (3.31,-0.3) .. (0,0) .. controls (3.31,0.3) and (6.95,1.4) .. (10.93,3.29)   ;
%Straight Lines [id:da4035800651514363] 
\draw    (420.67,174.33) -- (420.67,154.33) ;
%Straight Lines [id:da5395590510587387] 
\draw    (260.67,34.33) -- (280.67,34.33) ;
%Straight Lines [id:da7067108738586372] 
\draw [color={rgb, 255:red, 0; green, 0; blue, 0 }  ,draw opacity=1 ][line width=1.5]  [dash pattern={on 5.63pt off 4.5pt}]  (270.67,104.33) -- (340.67,164.33) ;
%Straight Lines [id:da6256717281249851] 
\draw [color={rgb, 255:red, 0; green, 0; blue, 0 }  ,draw opacity=1 ][line width=1.5]  [dash pattern={on 5.63pt off 4.5pt}]  (340.67,34.33) -- (420.67,104.33) ;
%Straight Lines [id:da7539315999944705] 
\draw  [dash pattern={on 0.84pt off 2.51pt}]  (270.67,104.33) -- (420.67,104.33) ;
%Straight Lines [id:da6533568262089124] 
\draw  [dash pattern={on 0.84pt off 2.51pt}]  (340.67,164.33) -- (340.67,34.33) ;
%Straight Lines [id:da56860436565969] 
\draw  [dash pattern={on 0.84pt off 2.51pt}]  (420.67,164.33) -- (420.67,34.33) ;
%Straight Lines [id:da07329107086131081] 
\draw  [dash pattern={on 0.84pt off 2.51pt}]  (184,165) -- (184,35) ;
%Straight Lines [id:da41174580183036524] 
\draw  [dash pattern={on 0.84pt off 2.51pt}]  (34,35) -- (184,35) ;
%Straight Lines [id:da8732766740516797] 
\draw  [dash pattern={on 0.84pt off 2.51pt}]  (270.67,34.33) -- (420.67,34.33) ;
%Straight Lines [id:da31466531094427874] 
\draw [line width=1.5]  [dash pattern={on 1.69pt off 2.76pt}]  (8.67,209) -- (58.67,209) ;
%Straight Lines [id:da9324388292425874] 
\draw [color={rgb, 255:red, 0; green, 0; blue, 0 }  ,draw opacity=1 ]   (112,209.67) -- (162,209.67) (122,205.67) -- (122,213.67)(132,205.67) -- (132,213.67)(142,205.67) -- (142,213.67)(152,205.67) -- (152,213.67) ;
%Straight Lines [id:da1565062661815475] 
\draw [color={rgb, 255:red, 0; green, 0; blue, 0 }  ,draw opacity=1 ][line width=1.5]  [dash pattern={on 5.63pt off 4.5pt}]  (223.33,210.33) -- (273.33,210.33) ;

% Text Node
\draw (41,7.4) node [anchor=north west][inner sep=0.75pt]    {$n/M$};
% Text Node
\draw (211,167.4) node [anchor=north west][inner sep=0.75pt]    {$k$};
% Text Node
\draw (157,177.4) node [anchor=north west][inner sep=0.75pt]  [font=\footnotesize]  {$M-1$};
% Text Node
\draw (14,97.4) node [anchor=north west][inner sep=0.75pt]  [font=\footnotesize]  {$u_{0}$};
% Text Node
\draw (393,176.73) node [anchor=north west][inner sep=0.75pt]  [font=\footnotesize]  {$M-1$};
% Text Node
\draw (222.33,22.73) node [anchor=north west][inner sep=0.75pt]  [font=\scriptsize]  {$\frac{M-1}{M}$};
% Text Node
\draw (448,167.73) node [anchor=north west][inner sep=0.75pt]    {$k$};
% Text Node
\draw (331.67,166.73) node [anchor=north west][inner sep=0.75pt]  [font=\footnotesize]  {$u_{0}$};
% Text Node
\draw (277.67,6.73) node [anchor=north west][inner sep=0.75pt]    {$n/M$};
% Text Node
\draw (250,96.73) node [anchor=north west][inner sep=0.75pt]  [font=\footnotesize]  {$u_{0}$};
% Text Node
\draw (69.67,201.4) node [anchor=north west][inner sep=0.75pt]  [font=\footnotesize]  {$x_{0}^{*}[ k]$};
% Text Node
\draw (173,202.07) node [anchor=north west][inner sep=0.75pt]  [font=\footnotesize]  {$s_{TJ}^{0}[ k]$};
% Text Node
\draw (283.33,196.73) node [anchor=north west][inner sep=0.75pt]  [font=\footnotesize]  {$\tilde{s}_{TJ}^{0}[ k] =s_{TJ}^{0}[ k] x_{0}^{*}[ k]$};
% Text Node
\draw (-13.67,24.07) node [anchor=north west][inner sep=0.75pt]  [font=\scriptsize]  {$\frac{M-1}{M}$};
\end{tikzpicture}